# Detecting cell and protein concentrations by the use of a thermal based sensor

Goossens J[1], Oudebrouckx G[1], Bormans S[1], Vandenryt T[1], Thoelen R[1]

[1]UHasselt, Institute for Materials Research (IMO-IMOMEC), Biomedical Device Engineering Research Group, Agoralaan, 3590 Diepenbeek, Belgium

*Running title: Development of a thermal based sensor*

Correspondence should be addressed to:
R. Thoelen, Tel: +32 (11) 26 88 29; Email: ronald.thoelen@uhasselt.be

## ABSTRACT

Biosensors are frequently used nowadays for the sake of their attractive capabilities. Because of their high accuracy and precision, they are more and more used in the medical sector. Natural receptors are mostly used, but their use have some specific drawbacks. Therefore, new read-out methods are being developed where there is no need for these receptors. Via a Transient Plane Source (TPS) sensor, the thermal properties of a fluid can be determined. These sensors can detect the capability of a fluid to absorb heat i.e. the thermal effusivity. By the way of monitoring this property, many potential bioprocesses can be monitored. The use of this promising technique was further developed in this research for later use of detecting cell growth and protein concentrations. Firstly, the thermal properties of growth medium and yeast cells were determined. Here, it became clear that the thermal properties change in different concentrations. Also, measurements were performed on protein concentration. No unambiguously results were obtained from these tests. But, the overall results from the use of this sensor are very promising, especially in the cell detection compartment. However, further research will tell about the applicability and sensitivity of this type of sensor.

## SAMENVATTING

Biosensoren worden de dag van vandaag veel gebruikt door hun aantrekkelijke eigenschappen. Door hun hoge nauwkeurigheid en precisie worden ze vaker gebruikt in de medische sector. Meestal wordt hier gebruikt gemaakt van natuurlijke receptoren, maar deze hebben zo hun eigen nadelen. Hierdoor worden nieuwe read-out methodes ontwikkeld waarbij geen nood is aan gebruik van receptoren. Via een Transient Plane Source (TPS) sensor kunnen thermische eigenschappen worden bepaald van een vloeistof. Deze sensoren bepalen hoe goed een vloeistof warmte kan opnemen m.a.w. de thermische effusiviteit. Door het monitoren van deze eigenschap, kunnen vele potentiele bioprocessen worden gemonitord. Deze veelbelovende techniek werd verder ontwikkeld in dit onderzoek om later te kunnen gebruiken voor de monitoring van cel groei en eiwit concentratie bepaling. Eerst worden de thermische eigenschappen van groeimedium en gistcelconcentraties bepaald. Hier werd duidelijk dat de eigenschappen veranderen in verschillende concentraties. Daarnaast worden ook metingen uitgevoerd op eiwit concentraties. Hierbij werden geen eenduidige resultaten bekomen. De resultaten van de sensor algemeen zijn veelbevolend, maar vooral in de detectie van cel concentraties. Echter, verder onderzoek zal uitwijzen over de toepasselijkheid en sensitiviteit van dit type sensor.

## INTRODUCTION

Biosensors are analytical tools that can detect the presence of a specific molecule or ligand. They are able to convert a biological response to an electrical signal [1]. These sensors consist of a bio-recognition element, which is specifically produced to detect the ligand, and a transducer, which can convert the biological signal to a measurable signal. In these kind of sensors, they use antibodies, phages, enzymes, etc. to indicate the presence of a target molecule [2].

Due to the outstanding accuracy and precise results from previous developed sensors, there is a high interest in the modern society for biosensors. The development of these sensors requires multidisciplinary knowledge in the field of chemistry, biology and engineering which makes it a complex research domain [3,4].

Unfortunately, there are several drawbacks when using natural receptors to detect a ligand. They can behave differently in diverse physical and chemical environments. They also display a limited-shelf life and it could be time consuming to obtain. By using a thermal- based sensor, some of these drawbacks could be overcome which also applies for other detection techniques like optical, micro gravimetrical, and others [5].

Transient Plane Source (TPS) sensors are being developed as a new thermal readout method. These sensors aren't expensive and are easy to use. The main focus of this type of sensor, is to determine the thermal properties of a sample. These properties can vary when it is heated or cooled down. This is due to the fact that energy is being supplied or removed from the sample. This change in thermal properties can be a temperature increase, phase transition, change of volume, etc [6]. This can be interesting when monitoring bioprocesses inside a fluid like for example the sedimentation of cells.

The sensor generates heat (Joule's heat) by a constant heating power which is produced by a current that is sourced through the sensor [7]. This heat is transferred to the sample by conduction and then evenly distributed inside the sensor by convection. Different samples react different to the thermal changes, this data can be used to determine the thermal conductivity ($\kappa$) and the thermal diffusivity ($\alpha$). This sensor operates as a heat source but also as a resistance thermometer [8,9].

The rate at which temperature changes take place is the thermal diffusivity, this indicates how quickly a material reacts to a change in temperature. The thermal conductivity is a measure of the ability of a material to conduct heat i.e. the rate of heat flow through a unit. The thermal conductivity is the reciprocal of the thermal resistance (r). By other means, this is a measurement of a temperature difference by which a material resists heat flow [6,10].

A key feature of the TPS sensor is the determination of this thermal effusivity of the fluid that is in contact with the sensor. Via this measurement, the heat flow from the sensor to the sample can be monitored. This gives an indication of the thermal properties.

Also, these TPS sensors are being developed to monitor cell growth. The accurate determination of the number of cells in a solution is important to monitor a bioprocess. Manual microscopic counting of cells could be a very labor-intensive job. A sensor specifically designed for this job, would be an immense ease in the lab [11].

If the TPS sensor would be covered in cells, the thermal effusivity at the sensor interface would decrease because the heat transfer from the sensor to the sample is being blocked. By this means, the number of cells present on top of the sensor could be monitored in function of the thermal resistance. This also applies for monitoring the sedimentation of cells or aggregates in a solution. These can also form a layer on top of the sensor and thereby increasing the thermal resistance [5].

P. Cornelis et al. used this heat transfer method to specifically detect E. coli by using SIP receptors. On top of the SIP layer are specific cavities for the E. coli to bind to [12]. They could selectively detect the target cells even if they were mixed with hundredfold excess of competitor cells. The sample of E. coli was mixed with other bacteria such as: Klebsiella pneumonia, Pseudomonas aeruginosa, Enterococcus cecorum, Staphylococcus epidermidis, etc. The amount of binding to the SIP layer was then monitored by measuring the thermal effusivity [13]. However, it will be challenging to detect the growth of those bacteria without using the associated receptor, such as the SIP layer in this case.

Promising results were already obtained by using this heat transfer method by this TPS sensor. Further research of this type of sensor will influence the applicability in the future. The goal in the end is to be able to monitor cell growth and DNA amplification. Hereby the question remains if a protocol could be standardized to measure these differences. Therefore, this TPS sensor has to be further developed and tested to fine-tune the measurements to the point that it can detect such little DNA concentration differences. Before testing with DNA samples, the goal is first to determine protein concentration in order to extrapolate this technique to hopefully measure DNA amounts in the future. Secondly, the protocol for detecting cell growth will be further developed to monitor the growth rate in the end. By the growth of yeast cells inside a medium, the thermal properties will change which would be detected by this thermal based sensor.

## MATERIALS AND METHODS

*Material and products* – Growth medium measurements were performed with YPD broth from VWR in a 50 g/l concentration. As yeast cells, normal dry yeast from Dr. Oetker was used. Bovine serum albumin (BSA) was diluted in a 0,1 g/ml concentration. The cleaning of tubings was performed with a 1% bleach dilution in milliQ.

*TPS Sensor* – The sensor used in this paper was the Flex v1.7. This sensor consists of a planar heating element in the form of a twisted copper wire. This is placed on a thin foil of a flexible polyimide PCB, which makes this sensor safe to use with various chemicals. The sensor performs as a transient plane source. This makes it possible to determine the thermal effusivity of the fluid but also the resistance on the surface itself. The metal inside is also used as a temperature sensor by using the temperature coefficient of resistance.

*Measuring protocol* – The source measure units (SMU) used in this study were a Keysight B2901A and a NI PXIe-4139. These devices have the ability of supplying power to the sensor and measuring at the same time. It creates a block wave pulse and the heating time, heating temperature, sample frequency and cooling time could be installed. This cooling time provides a short window for the sensor to cool down to room temperature. To check if the sensor worked properly, the initial resistance was always checked before performing any measurements.

*Flow cell measurement* - The sensor was placed inside a flow cell where it's in contact with the sample on one side and covered by a piece of glass on the other side to ensure a tight fit. Therefore, the sample was always visible, which makes it possible to identify possible air bubbles. This chamber was always fully filled and therefore always free of air bubbles which could otherwise influence the measurement. This chamber can hold approximately 3 ml of fluid. The housing has an in- and outlet for convenience of measuring different samples inside this chamber (Figure 1). These measurements were performed with the PXIe with a heating time of 1 s, heating power of 0.5 W, sample frequency of 500 Hz and a cooling time of 60 s. For the ease of use, a syringe pump was used to automatically fill and empty the flow cell with different fluids. This made it possible to perform multiple measurements one after another. In this research the Laboratory Syringe Pump LSPone by Advanced Microfluidics was used. It has 8 different channels and has a maximum pumping volume of 1000 µl.

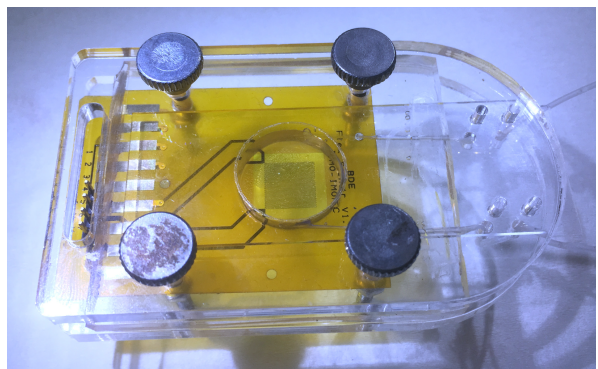

**Figure 1: Flow cell measurement setup with the Flex V1.7.** Measuring setup where the flow cell is in contact with the sensor and a layer of glass on the other side.

*Open well measurement* – An open well setup has the ability to be cleaned easily. The Flex V1.7 sensor was taped between 2 pieces of PMMA and screwed together to ensure an even tighter fit. The pieces of PMMA were laser cut to the exact same dimensions as the Flex V1.7. A hole was cut over the heating element where a sample could be inserted (Figure 2). Air is a good thermal insulator and therefore a hole was made at the back of the heating element. This ensures all the heat is transferred to the sample. The open well was covered by a piece of glass during the measurements to avoid evaporation. The well was manually filled with 1,5 ml of the sample with a micropipette. These measurements were performed with the PXIe or Keysight with a heating time of 1 s, heating power of 0.5 W, sample frequency of 300 Hz and a cooling time of 60 s.

*DIP measurement* – A DIP sensors works on the same principle of TPS. This type of sensor has no need for a flow cell or open well but can just be dipped inside a fluid. A Pasteur pipette is used as casing for the wiring. A tiny heating element is exposed at the tip of the pipette (Figure 3). The sensor was fixated on a tripod and dipped just as deep each measurement. DIP measurements were performed in a falcon tube of 15 ml and filled with 10 ml of fluid. These measurements were performed with the PXIe with a heating time of 1 s, heating power of 0.2 W, sample frequency of 500 Hz and a cooling time of 60 s.

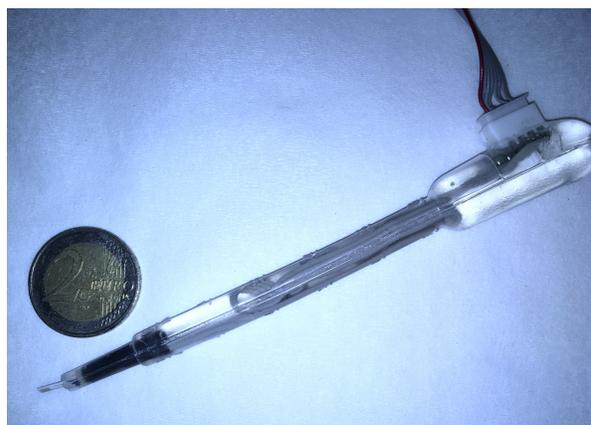

**Figure 3: DIP sensor measurement setup.** The tip of the sensor works on the same principle as the Flex V1.7. The heating element is situated at the tip of the Pasteur pipette. €2 coin is placed as size reference.

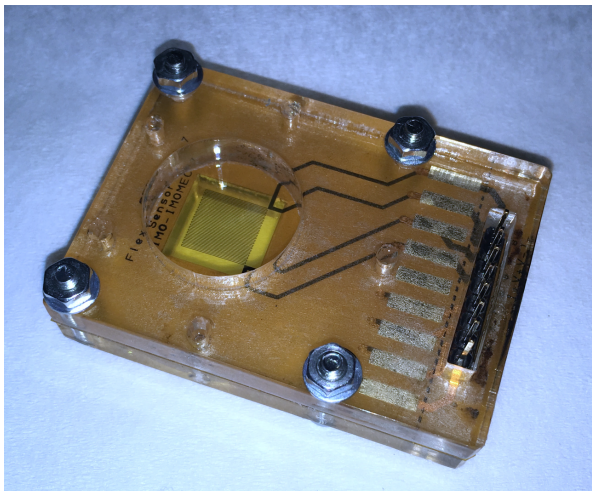

**Figure 2: Open well measurement setup using Flex V1.7.** The sensor is sandwiched by 2 pieces of PMMA. To ensure a watertight fit, double sided tape was added. The open well holds max 1,8 ml of fluid. The open well is covered by a piece of glass when measuring.

*Thermal properties measurement* – When the temperature increases inside the sensor, it will dissipate the heat to the sample. The thermal effusivity is the ability of a material to absorb heat. This thermal effusivity (e) is given by the square root of the thermal conductivity ($\lambda$) of a material with density ($\rho$) and heat capacity (c). (Eq. 1)

$$e = \sqrt{\lambda \rho c} \qquad (1)$$

When supplying the sensor with a block wave current pulse (Figure 4c), the electric resistance increases (Figure 4d). The electric resistance of the sensor raises because of the given current. When a layer of cells sediments on top of the sensor (Figure 4a+b), the resistance of the heat transfer raises even more (Figure 4d). Thus, the higher the electrical resistance, the lower the thermal conductivity and therefore lower effusivity. How steeper the slope of the resistance, how lower its capacity to conduct the heat transfer. This slope will be used in the following paper to indicate the thermal effusivity. When looking at the slope at the end of the curve, the heat transfer is seen inside the sample. When looking at the slope at the beginning of the curve, the heat transfer is seen on top of the sensor.

*Data processing* – All the data gathered by using these sensors were uploaded onto a local network, where it is available for all researchers in this group. Via this site, all TPS analysis were made. The data was also automatically plotted, and these graphs and plots were used in the results.

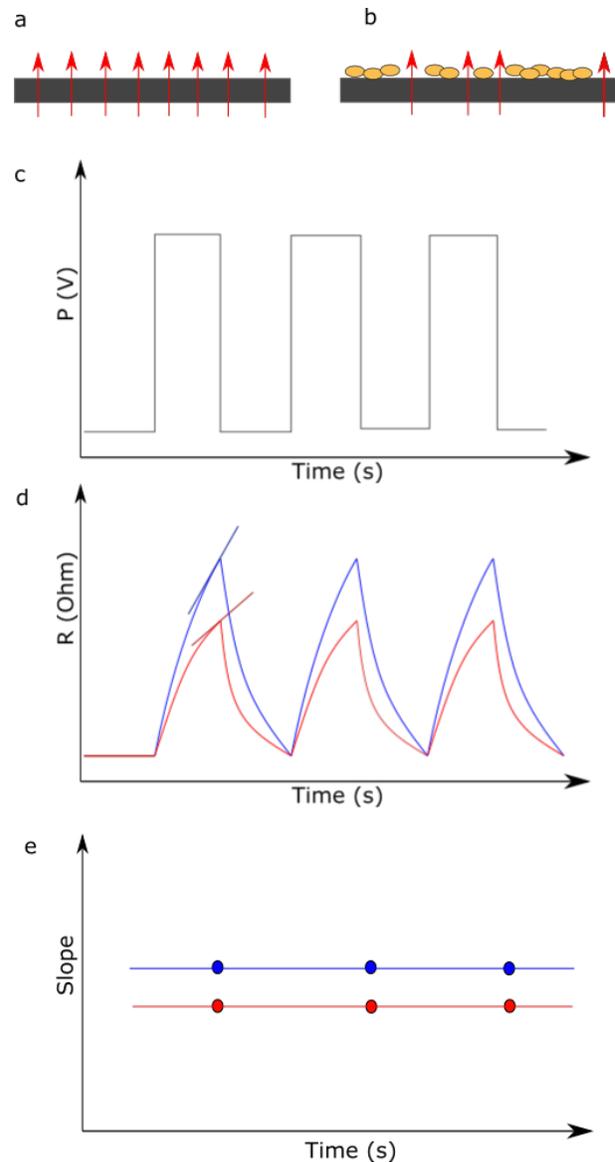

**Figure 4: Response of a TPS sensor to determine thermal properties of a sample.** A+B) Schematic representation of sedimentation of cells on the sensor to visualize the blocking of heat transfer. Heat is represented by the red arrows. C) Visualization of the power supply by a SMU in a block wave pulse. D) When the electric resistance raises, the thermal effusivity drops, which indicates the presence of sedimented cells. Likewise, the blue curve indicates that in the case of b) the heat is blocked and the electric resistance increases. Case a) is given by the red curve. E) The slopes of the resistance in d) are given, here its seen that the slope of the blue curve is higher than the slope of the red one. The effusivity of the red curve is higher.

**RESULTS**

*Determination of reference value* – For all measurements, the thermal effusivity of water was considered as the reference value. This also gives an indication if the sensors worked properly. Here, the goal is to have a stable slope signal over time.

*Measurement of thermal properties of yeast cells in flow cell* – Before measurements could be performed on growing cells. First measurements were executed to detect different cell concentrations with the flex V1.7 inside the flow cell. Via this TPS method, we have the ability to detect a difference in effusivity. Also, a distinction can be made between the resistance on the surface and the resistance of the fluid inside the flow cell. When the resistance is plotted at the beginning of the slope, the resistance is seen at the surface of the sensor. This can clearly be seen when yeast is dissolved in milliQ and flushed inside the sensor. The resistance of the heat transfer increases when the cells sediment over time on top of the sensor (Figure 5). In contrast, the thermal properties of the fluid itself (in this case milliQ) does not change that significant. This is seen when looking at the end of the slopes (Figure 6). When flushing different yeast concentrations inside the flow cell, a problem arises. After 20 minutes the yeast has been sedimented on top of the sensor. Even after flushing multiple times with milliQ, the yeast always stays a little inside the flow cell.

*Measurement of thermal properties of yeast cells with DIP sensor* – A DIP sensor is used to dip the sensor inside the fluid and hereby measuring the thermal properties of the sample. A big difference in using this type of sensor, is that it is vertically placed in the sample. In contrast to the flow cell where the fluid is pumped over the sensor's surface. When using this sensor, it became clear that it is hard to get a stable signal (Figure 7). The tip of the sensor was always cleaned with ethanol and a cotton bud. Even then the water measurements weren't stable as using a Flex V1.7.

*Measurement of thermal properties of yeast cells in open well* – An open well enables to measure horizontally and can be easily cleaned. The Flex V1.7 was used to perform these measurements. Different concentrations were loaded into the open well and the thermal effusivity was measured. The yeast cells sedimented on top of the sensor over time. Here, the slopes of different concentrations (20 mg/ml and 10 mg/ml) indicate a difference of thermal properties (Figure 8). The data obtained from this measurement was extremely clear, probably because the well could be thoroughly cleaned between different measurements.

*Thermal properties of growth medium* – YPD medium culture was dissolved in milliQ. The thermal effusivity of different concentrations was measured with the flex V1.7 inside the flow cell. As seen in the graphs, the thermal resistance of the fluid increases with increasing concentrations (Figure 9). Water flushes were performed in between different YPD concentrations as stability check.

a)

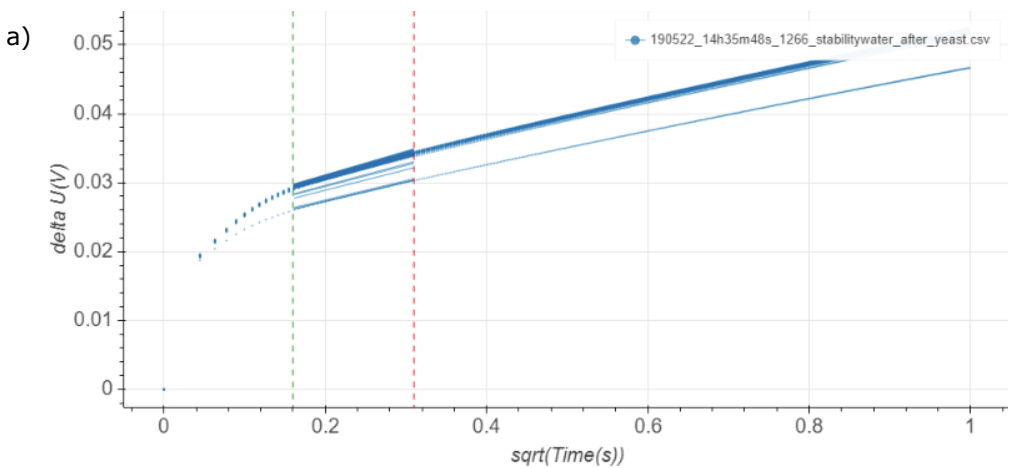

**Figure 5: Resistance at the height of the surface.** Water has a stable resistance. A) The green and red dotted lines indicate the interval were the slopes were plotted. B) When yeast (20 mg/ml in milliQ) is flushed inside the flow cell the cells sediment on top of the sensor and consequently prevents the heat transfer. Measurements were performed for at least 20 minutes (n=4).

b)

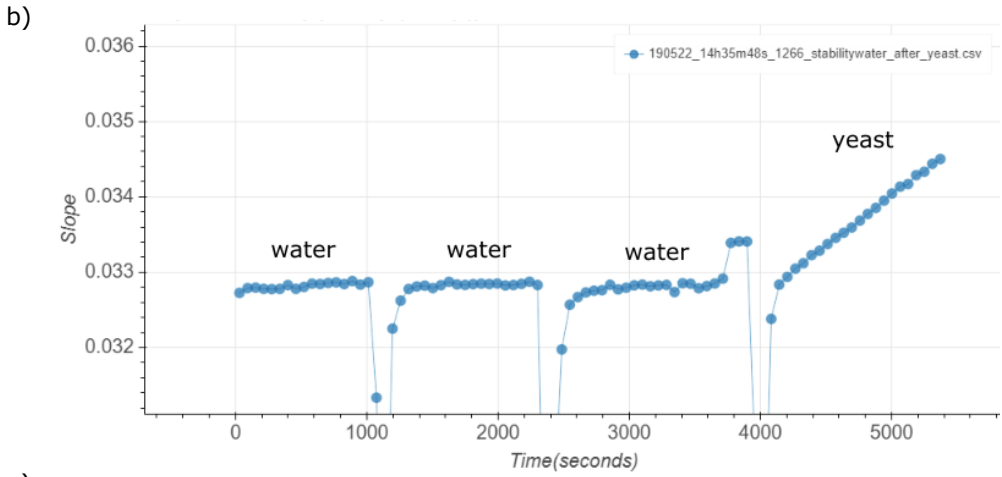

a)

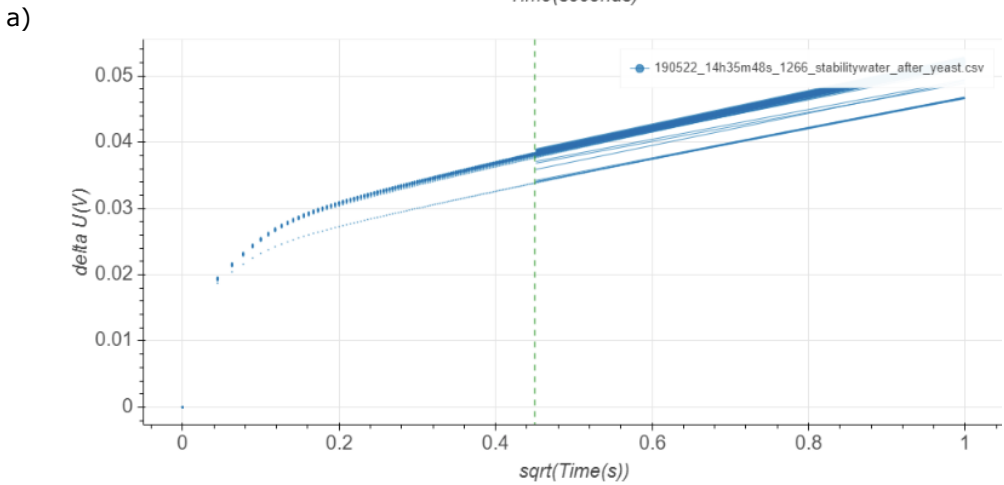

**Figure 6: Resistance at the height of the fluid.** Data from the same measurement as in figure 5. A) Here the slopes are plotted at the end of the curve, which gives an indication of the resistance of the fluid. B) The presence of yeast (20 mg/ml in milliQ) does not influence the thermal property drastically. Measurements were performed for at least 20 minutes (n=4).

b)

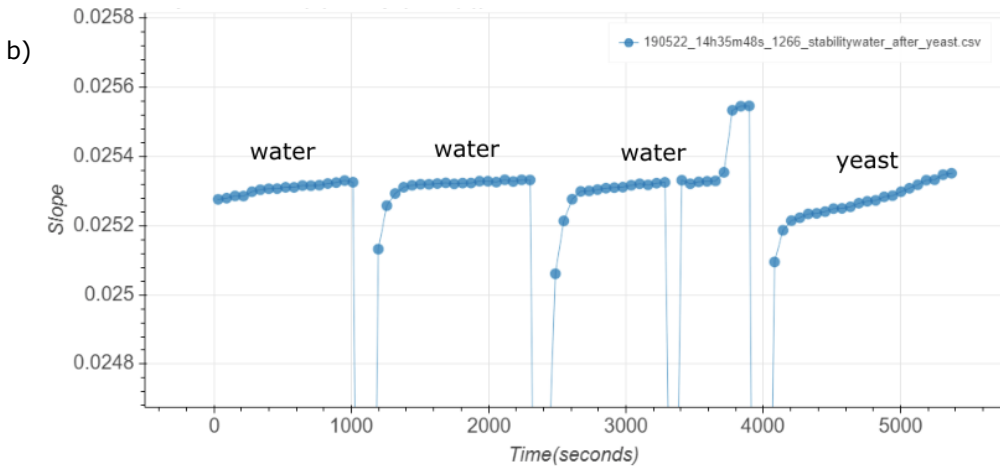

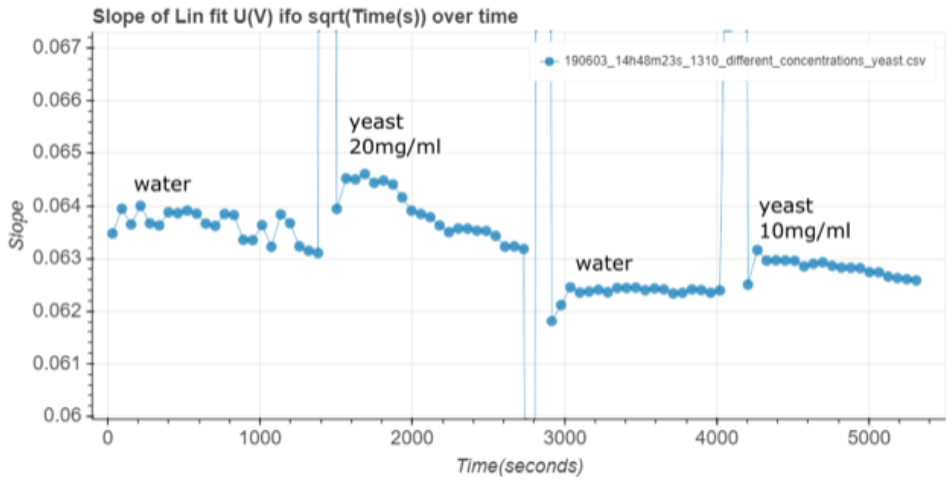

**Figure 7: The use of a DIP sensor to detect different yeast concentrations.** Unstable signal when using this type of sensor. No conclusions can be made when using this DIP sensor as long as the reference of water isn't stable. Measurements were performed for 20 minutes (n=4).

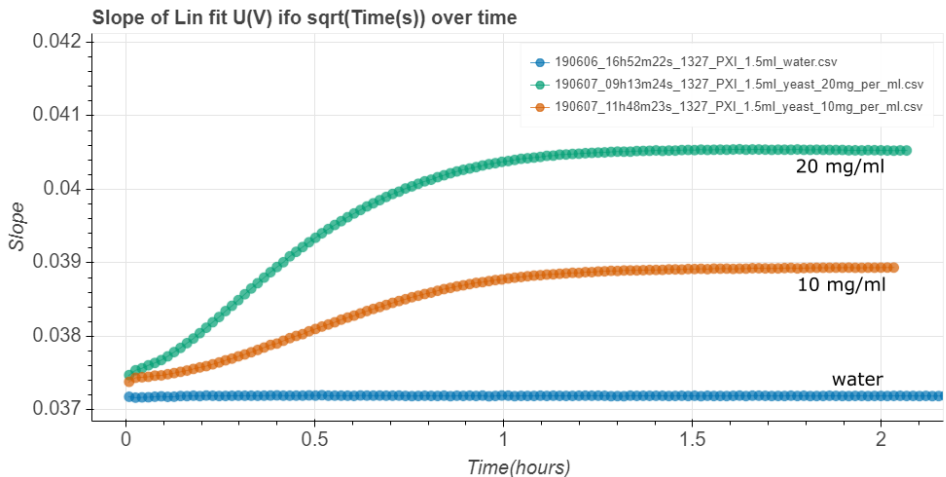

**Figure 8: Detection of different concentrations of yeast cells in milliQ.** 2 different concentrations of yeast cells were measured (green: 20 mg/ml, orange: 10 mg/ml) and water as reference. Measurements were performed for 2 hours (n=3).

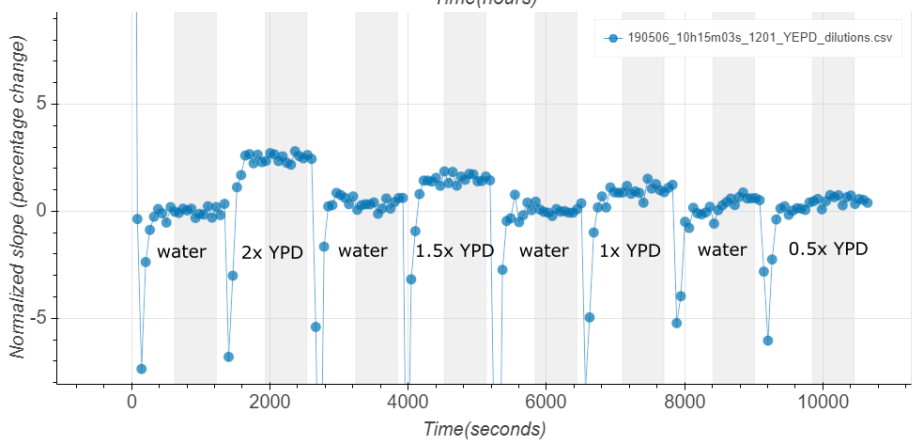

**Figure 9: Thermal resistance of different YPD concentrations in reference to water.** It's clearly seen that different concentration can be detected by this method. Each measurement was performed for 20 minutes (n=8). The flow cell was always flushed with water between measurement and to exclude sedimentation on the sensor, the effusivity of water was also measured. Slope is normalized to the reference value of water.

*Measurement of protein concentration* – The first measurement of a protein concentration was a BSA solution of 0,1 g/ml. The problem with using this protein is that it forms a persistent layer on top of the sensor and inside the flow cell and tubing. This BSA prevents the heat transfer from the sensor to the sample.  This gives a wrong indication of the thermal effusivity of the fluid. When measuring water as reference afterwards, it is clearly seen that the sensor is contaminated (Figure 10). By other means, this layer has to be removed from the sensor in order measure the thermal effusivity accurately. Even if the tubing and sensor were flushed with water and ethanol, the data still drifted towards a higher resistance. This gives an indication that the BSA remnants inside the tubing and sensor were persistent enough to adhere on the inside. When the sensor and tubing were cleaned with a 0.1% bleach solution, the first few data points seemed as normal. But, when the sensor kept running, the data points still drifted towards a higher thermal resistance. Even after a thorough clean of the flow cell and a manual flush of 20 ml bleach solution with high flowrate, the datapoints  still drifted away (Figure 11). This indicates that BSA still remained inside the tubing and was piling up on top of the sensor while measuring.

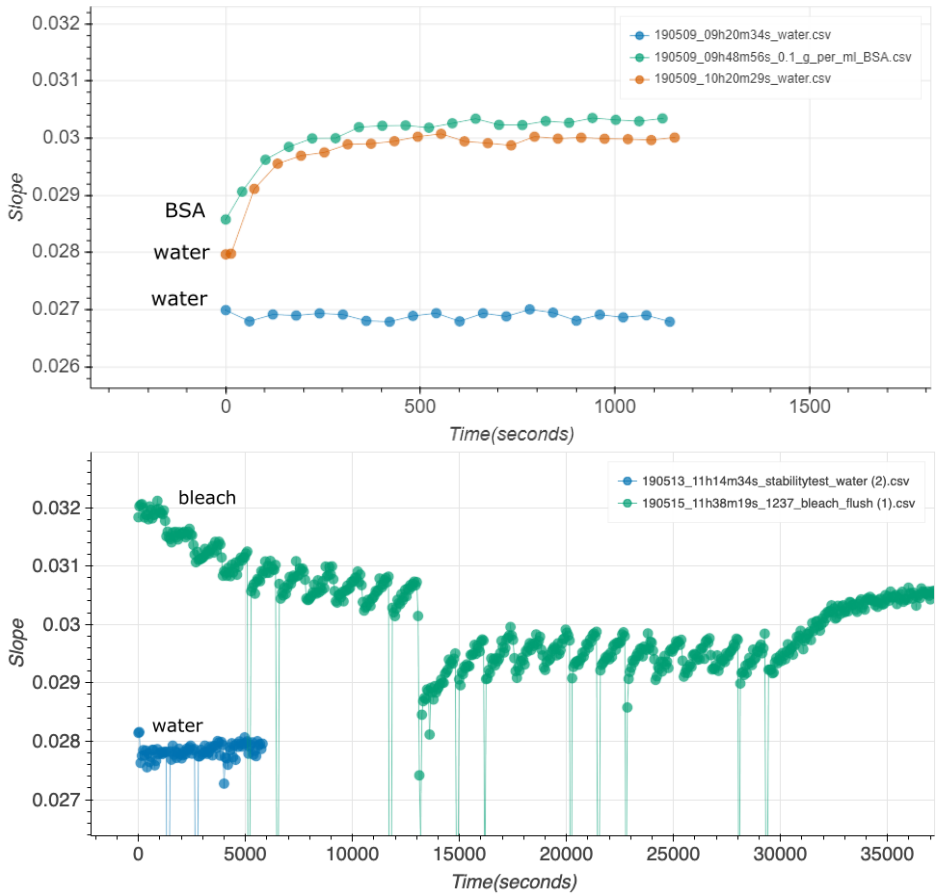

**Figure 10: Increasing resistance of BSA.** Water (blue curve) was measured and gave a stable signal. Afterwards BSA (green curve) was measured and the slopes increased significantly. When measuring water again, the slopes also increased drastically. This suggests a layer formation of BSA on top of the sensor. Measurements were performed for 20 minutes (n=3).

**Figure 11: Bleach cleaning of tubing.** When using bleach as a sample inside the sensor, the tubing was cleaned at the same time. Consequently, the BSA is removed from the tubing but stay inside the flow cell. When the flow cell is cleaned and rinsed manually with a bleach solution, a decrease in resistance is seen. Afterwards, the data still drifted towards a higher resistance by BSA remnants coming from the tubing.

## DISCUSSION

When using this TPS sensor, it became clear it is very precise but also super sensitive. Each measurement had to be performed exactly the same otherwise the measurement would fail. This gave a lot of reproducibility issues. The setup was constantly fine-tuned to get a stable signal for each measurement. These issues were tackled one by one. In the end, the setup could trustingly perform the measurement. Measurements were then performed multiple times to ensure its validity.

When fitting the slopes over time, the interval had to be set manually. Different fits provided different results. By this means, the results can be fine-tuned to the outcome the researcher would like. Therefore, the researchers had to be consistent when plotting the interval. Further work is being invested to develop a protocol where the measurements will be interpreted the same way each time.

The manual filling of the flow cell was a very labor-intensive work. Therefore, the Syringe Pump LSPONE was used, but this use arises other problems. Sometime the pump would suck up air, which results in air bubbles when pumping in the flow cell. These could influence the measurement. Also, the tubings were always primed with the sample. This would make sure no contamination found place inside the tubing. But this became doubtful after the contamination of BSA.

The BSA solution that was used in this study was a 0,1 g/ml dilution. By other means a 1,5 M concentration, which could have been too high. This could explain why the BSA stayed inside the tubing even after cleaning with bleach and ethanol. A lot of solutions were thought of to remove the BSA remnants. Other solvents e.g. Tween-20, SDS, etc. could remove the BSA from the substrate [14]. But this would arise another problem, because these solvents would also form an adhesive layer [15,16]. To this date, no solution was found to remove the BSA efficiently to perform consecutive measurements.

The results of the DIP sensor weren't as expected. The sensor couldn't perform a stable signal. This could be caused by the fact that the sensor was vertically placed inside the fluid. An important factor of detecting cell suspensions was the sedimentation which couldn't be detected when the sensor was placed vertically. Also, a cell suspension is not a homogenous fluid, which is maybe too difficult to detect with this tiny sensor. For measuring with this sensor, the sensor was suspended at a predefined height and dipped in a falcon tube of 15 ml which was filled with 10 ml of fluid. Even with this conformation, the water measurement didn't get stable.

Open well measurements were performed for the ease of cleaning. The first design was constructed with 2 caps of a falcon tube and a Flex V1.7. A hole was made in the cap, so the heating element was in contact with the sample. The cap on the bottom made sure the sensor was in contact with air on the other side. This contraption was fixed in place with a weight on top (Supporting information: Figure 12). With this primitive setup promising results were obtained (Supporting information: Figure 13). This design was later fine-tuned to a stable design of PMMA and even better results were obtained.

The Flex V1.7 always had to be handled with care. Especially the point where the sensor was soldered to the connector was extremely fragile. Sometimes this connection broke, and this sensor was then unable to perform another measurement. Therefore, multiple sensors were used during this research. Taken in to account that none of these sensors have the exact same initial resistance, results had to compared relative to each other. Despite the variation of different sensors, the results were almost always the same.

Different SMU's were used in this research. The advantage of using the PXIe is that the voltage range was 1-6 V. While the Keysight has a voltage range of 1-20 V. This advantage ensures a more stable signal, in contrast to the noisier data obtained by the Keysight. When comparing the Keysight to the PXIe, the difference in noise can be easily spotted (Supportive information: Figure 14). Despite the level of noise in the data, the results show the same outcome.

## CONCLUSION

In conclusion, this research shows the power of using a TPS sensor. This sensor can detect different concentrations growth medium and yeast cells. But it's thought the limitation aren't reached yet. This promising technique can potentially be further developed to detect cell growth. In this research, no protein concentrations could be detected but it is not considered impossible.

*Acknowledgements* – The authors are grateful for the technical and theoretical support provided by the staff of the Institute for Material Research (IMO-IMOMEC). Special thanks to the personal of the research group of Biomedical Device Engineering for the personal advice and support in performing this research.

**SUPPORTING INFORMATION**

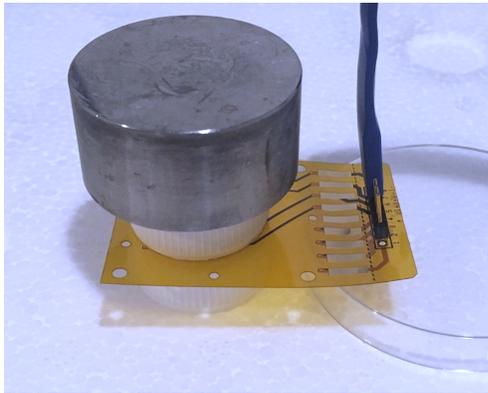

**Figure 12: Primitive open well measurement setup.**
Open well construction with 2 caps of a falcon tube and a Flex V1.7. The weight on top ensures that the setup stays in place. Petri dish under cabling ensures that the sensor does not bend too much.

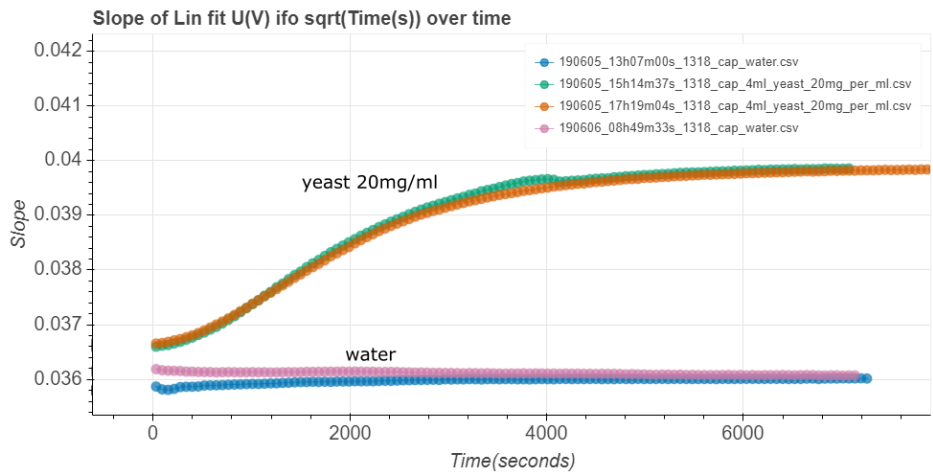

**Figure 13: Stable results obtained by using an open well setup.**
Even with the use of a primitive setup, this type of measurement obtained promising results. This gave an indication that the use of an open well has some serious advantages.

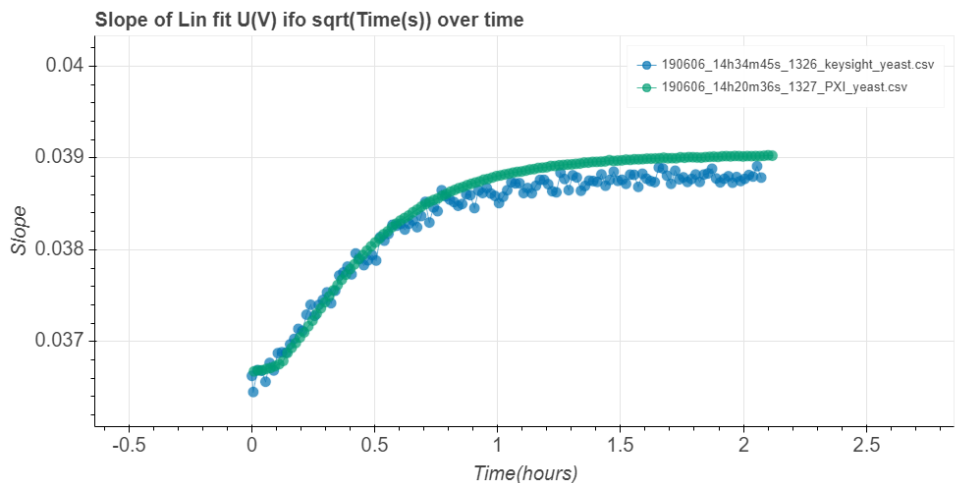

**Figure 14: The use of different Source Measure Units**
When using the PXIe, the data shows a clearer result. In contrast to the noisier data obtained by the Keysight. Despite the level of noise in the data, the results show the same outcome.